\documentclass{iau}
\usepackage{graphicx}
\usepackage{natbib}
\usepackage{amsmath}

\def\apj{{\it ApJ}}

\def\apjs{{\it ApJ. S}}

\def\apjl{ApJL}
\def\mnras{MNRAS}

\newif\ifAMStwofonts
\AMStwofontstrue

\def\vnabla{\bm{\nabla}}

\def\gsim{~\rlap{$>$}{\lower 1.0ex\hbox{$\sim$}}}

\def\simpropto{\lower.2ex\hbox{$\; \buildrel \propto \over \sim \;$}}
\def\ltsim{\lower.5ex\hbox{$\; \buildrel < \over \sim \;$}}
\def\gtsim{\lower.5ex\hbox{$\; \buildrel > \over \sim \;$}}
\def\ltsim{\lower.5ex\hbox{$\; \buildrel < \over \sim \;$}}
\def\gtsim{\lower.5ex\hbox{$\; \buildrel > \over \sim \;$}}

\def\vg{v_{_{\rm g}}}

\def\vnabla{{\bf \nabla}}

\def\kms{\mbox{km\,s$^{-1}$}}

\def\prd{Physical Review D.}

\def\astjo{Astronomical Journal}

\def\dd{\,{\rm d}}

\def\nb{{n_{_{\rm b}}}}

\def\kms{\ {\rm km\,s^{-1}}}

\def\dd{{\rm d}}
\def\nb{{\bar n}}
\def\ln{{\rm ln}}

\def\pmb#1{\setbox0=\hbox{#1}%
\kern-.025em\copy0\kern-\wd0
\kern.05em\copy0\kern-\wd0
\kern-.025em\raise.0433em\box0}

\def\vv{\pmb{$v$}}

\def\vs{\pmb{$s$}}

\def\vg{\pmb{$g$}}

\def\iras {{\it IRAS~}}

\def\etal{{\it et al.\ }}

\def\simlt{\lower.5ex\hbox{$\; \buildrel < \over \sim \;$}}
\def\simgt{\lower.5ex\hbox{$\; \buildrel > \over \sim \;$}}

\def\vnabla{\pmb{$\nabla$}}
\def\vitf{v_{_{\rm itf} }}
\def\vvitf{\bf \vitf}

\newcommand{\beq}{\begin{equation}}
\newcommand{\eeq}{\end{equation}}
\def\beqa{\begin{eqnarray}}
\def\eeqa{\end{eqnarray}}
\def\fixit#1{}

\def\dd{{\rm d}}
\def\apj{ApJ}

\def\apjs{ApJ. S}

\def\apjl{ApJL}
\def\mnras{MNRAS}

\title[Large Scale Flows]{Re-examination of Large Scale Structure  \\  \& Cosmic Flows }
\author{Marc Davis$^1$ \& Adi Nusser$^2$}
\affiliation{$^1$Departments of Astronomy \&  Physics,  University of California at Berkeley, CA 94720 
      \\ email: {\it mdavis@berkeley.edu} \\[\affilskip]    
      $^2$  Physics Department and the Asher Space Science Institute-Technion, \\ Haifa 32000, Israel
     \\ email: {\it adi@physics.technion.ac.il} }

\pubyear{2014}
\volume{308}  %% insert here IAU Symposium No.
\pagerange{xx-yy}
\jname{The Zeldovich Universe: Genesis and Growth of the Cosmic Web }
\editors{Rien van de Weygaert, Sergei Shandarin, Enn Saar \& Jaan Einasto}

\begin{document}
\maketitle

\begin{abstract}
{Comparison of  galaxy flows with those predicted from the  local galaxy distribution ended as an active field after two analyses came to vastly different conclusions 25 years ago, but that was due to faulty data.  All the old results are therefore suspect.  With new data collected in the last several years,   the problem deserves another look.  
 For this we analyze  the  gravity field inferred from the enormous data set derived from the 2MASS  collection of galaxies \citep{fhuch}, and compare it to the velocity field derived from the well calibrated SFI++   Tully-Fisher catalog \citep{spring07}.   Using the "Inverse Method" to minimize Malmquist biases, within 10,000 km/s the gravity field is seen to predict the velocity field \citep{Davis11}  to remarkable consistency.   This is a beautiful demonstration of linear perturbation theory  and is fully consistent  with standard values of the cosmological variables.  }
\end{abstract} 

\firstsection
\section{Comparison  of Observed  Velocity Field with Gravitational Field}

This is a conference proceeding where I summarize several recent publications on peculiar velocities.  In particular  the brief discussion is based on
\cite{Davis11} (hereafter D11), \cite{ND94} (hereafter ND94), and \cite{DNW96} (hereafter DNW96).  Interested parties will find complete references therein.
 
The analysis of \cite{Davis11} fits the  peculiar velocity field given by the SFI++  Tully-Fisher whole sky sample of 2830  galaxies with redshifts $cz < 10,000$ km/s \citep{spring07} to a set of orthogonal polynomials by means of an inverse Tully-Fisher (ITF)  procedure. The  peculiar velocity field derived from this sample is then compared to the gravity field from the  largest whole sky redshift survey, the 2MRS  survey \citep{fhuch}. This catalog is   $K$ band selected 2MASS galaxies and has been extended to 43,500 galaxies to $K \le 11.75$  and 
$ |b| > 5^\circ$ or $ |b| > 10^\circ$  
near the galactic center. In our lifetime, the redshift catalog and derived gravity field is unlikely to improve enough to bother, since it is not the limiting noise. For improvements in the future, one should work on enlarging the TF data.

 Peculiar velocities are unique in that they
provide explicit information on the  three dimensional mass
distribution, and measure mass on scales of $10-60 h^{-1}$ Mpc, a scale 
untouched by alternative methods.  
%Local  peculiar velocity data  is, in principle,  affluent in 
%cosmological information.
%Power spectra  and correlation functions 
%could be derived from the data by direct calculation or  by maximum likelihood techniques \citep[e.g.][]{gd89, jk95, freu99,jusf00,brid01,abate09}.  Direct calculation of low order moments of the flow, such as the bulk
%motion and the shear could also analyzed within the framework of cosmological models 
%\citep[e.g.][]{feldwh10}.  Some authors claim the bulk flow of mixed catalogs of galaxies argues there are problems with 
%$\Lambda CDM$ \citep{watkins09}, but recent results \citep{nuss11} show that the SFI++ catalog by itself has a large scale bulk flow that is consistent with  $\Lambda CDM$, and this analysis has smaller error bars. 
%Other, perhaps more ambitious,  applications could involve an assessment of the 
%statistical nature of the initial cosmological large scale fluctuations, i.e. whether gaussian or otherwise 
%\citep{nd93,  berj95}.  All these analyses could be performed with peculiar velocity measurements alone.  will him and
Here we will be concerned with a comparison of the observed peculiar velocities on the one  hand and the velocities derived from  the fluctuations in the galaxy distribution 
on the other. The basic physical principle behind this comparison is simple.
The large scale flows are almost certainly the result of the process
of gravitational instability   with overdense regions attracting material, and 
underdense  regions repelling material.  Initial conditions in the early universe
might have been somewhat chaotic, so that the original peculiar
velocity field (i.e.  deviations from Hubble flow) was uncorrelated
with the mass distribution, or even contained vorticity.  But those
components of the velocity field which are not coherent with the
density fluctuations will adiabatically decay as the Universe expands,
and so at late times one expects the velocity field to be aligned with
the gravity field, at least in the limit of small amplitude
fluctuations \citep{Peeb80, n91}.
In the linear regime, this relation implies a simple proportionality between 
the gravity field {\bf g} and the velocity field ${\bf v}_{\rm g}$,
namely   ${\bf v}_{\rm g} \propto {\bf g}~ t$ where 
the only possible time $t$ is the Hubble time.  The exact expression
depends on the mean cosmological density 
parameter $\Omega$  and is given by \cite{Peeb80},
\begin{equation}
\label{eq:vg}
{\bf v}_{\rm g}(r) = {\frac{2 f(\Omega)}{ 3 H_0 \Omega}} {\bf g}({\bf r}) \ . 
\end{equation} 
%where $f(\Omega)$ is the linear growth rate, \citep{lind05}. 
  Given complete knowledge of the mass fluctuation field $\delta_\rho({\bf r})$
over all space,  the gravity field ${\bf g(r) } $ is
\begin{equation}
\label{eq:grho}
 {\bf g}({\bf r}) =  G\bar{\rho} \int d^3 {\bf r'} \delta_\rho({\bf r'})
\frac{{\bf  r' }-{\bf  r}} { |{\bf r'}-{\bf r}|^3} \ , 
\end{equation}
where $\bar\rho$ is the mean mass density of the Universe.  
If the galaxy distribution at least
approximately traces the mass on large scale, with linear bias $b$
between the galaxy fluctuations $\delta_G$ and the mass fluctuations
(i.e. $\delta_g = b \delta_\rho$),
then from (\ref{eq:vg}) and (\ref{eq:grho}) we have  
\begin{equation}
\label{eq:vrho}
{\bf v}_{\rm g}(r) = {H_0 \beta \over 4 \pi {\bar n}}
 \sum_{i} {1\over\phi(r_i)} {\bf  r_{\rm i} -r  \over | r_{\rm i} -r|^3 }
+ {{H_0 \beta }\over 3}{\bf r}\ , 
\end{equation}
where $\bar n$ is the true mean galaxy density in the sample, $\beta \equiv f(\Omega)/b$ with $f\approx \Omega^{0.55}$ 
the linear growth factor \citep{lind05},  and where we have replaced the integral over space
with a sum over the galaxies in a catalog, with radial selection
function $\phi(r)$\footnote{$\phi(r)$ is
defined as the fraction of the luminosity distribution function observable
at distance $r$ for a given flux limit; see \citep[e.g.][]{y91}.}.
 The second term is 
for the uniform component of the galaxy distribution and  would exactly
cancel the first term in the absence of clustering within the survey volume.  Note that the result is insensitive to the value of $H_0$, as the right hand side has units of velocity.  We shall henceforth quote
all distances in units of $\kms$. 
The sum in equation  (\ref{eq:vrho})  is to be computed in real space, whereas the galaxy catalog exists in redshift space. As we shall see in \S\ref{sec:zdist}, the modified equation,  which 
includes redshift distortions, maintains a dependence on $\Omega$ and $b$ through the 
parameter $\beta$.
Therefore,   a    comparison of the measured velocities of galaxies  to the
predicted velocities, ${\bf v}_{\rm g}(r)$, gives us a measure of $\beta$.
Further, a detailed comparison of the flow patterns addresses 
fundamental questions regarding the way galaxies trace mass on large scales and 
the validity of gravitational instability theory. 

\section{Methods}
In this section we outline our method described in ND94, ND95 and DNW96  for deriving the smooth peculiar velocities of galaxies from
an observed distribution of galaxies in redshift space and, independently,
from a sample of spiral galaxies with measured circular velocities 
$\eta$ and apparent magnitudes $m$.  

%\subsection{Peculiar Velocities from the Distribution of Galaxies in Redshift Space}
\label{sec:zdist}
%There are several   methods
%for generating peculiar velocities from redshift surveys, 
%using linear  \citep[e.g.][]{Fisher95b}  and non-linear relations 
 %\citep[e.g.][]{Peeb80,cc98,NussBranch,Frisch, Ensslin09}
Here we restrict ourselves to large scales where linear-theory is applicable.
We will use the  method of ND94  for reconstructing velocities from the 2MRS. This method is 
particularly convenient, as it is easy to implement, fast,
and requires no iterations. Most importantly, this redshift space
analysis closely parallels the ITF estimate described below. 
We next present  a very brief summary of the
methodology.

We follow the notation of DNW96. The comoving redshift space coordinate and 
the comoving peculiar velocity relative to the Local Group (LG) are, respectively, denoted by 
 $\vs $ (i.e. $s= cz/H_0$) and $\vv(\vs)$.
To first order, the peculiar velocity is irrotational
in redshift space \citep{chodnuss} and can be expressed  as $\vv_g(\vs)=-\vnabla\Phi(\vs)$ where  $\Phi(\vs)$ is a potential function.
As an estimate of the fluctuations in the fractional density field $\delta_0(\vs)$ traced by the discrete distribution 
of galaxies in redshift space  we consider,
\begin{equation}
\delta_0(\vs)= 
{1\over {(2\pi)^{3/2}\nb \sigma^3}}\sum_i {{w(L_{0i})}\over {\phi(s_i)}} 
\exp\left[-{{\left(\vs - \vs_i\right)^2}\over
{2\sigma^2}}\right] -1 \quad . 
\label{eq:deldef}
\end{equation}
where $\bar n=\sum_i w(L_{0i})/\phi(s_i)$ and $ w$ weighs  each   galaxy according to its estimated luminosity, $L_{0i}$. 
The 2MRS density field is here smoothed by a gaussian window 
with a redshift independent width, 
$\sigma=350\kms$. This is in contrast to DNW96 where the \iras 
density was smoothed with a width proportional to the mean particle separation.
The reason for adopting a constant smoothing for 2MRS is its dense sampling which is nearly four time 
higher than  \iras.
We emphasize that the coordinates ${\bf s}$ are in {\it
observed redshift} space, expanded in a galactic reference frame.  The only
correction from pure redshift space coordinates is the
collapse of the fingers
of god of the known rich clusters prior to the redshift space smoothing
(Yahil \etal 1991).
Weighting  the galaxies in equation (\ref{eq:deldef})  by
the selection function and luminosities evaluated at their redshifts rather than the
actual (unknown) distances yields a biased estimate for
the density field.  This bias   gives rise to Kaiser's rocket effect \citep{kais87}.  

To construct the density field,  equation \ref{eq:deldef}, we volume limit the 2MRS sample to 3000 km/s, so that $\phi{(s<3000)}= 1$, resulting in $\phi{(s=10000)}=0.27$ \citep{w09}.  In practice, this means we delete galaxies from the 2MRS sample fainter than $M_* + 2$.  Galaxies at 10,000 km/s therefore have $1/\phi = 3.7$ times the weight of foreground galaxies in the generation of the velocity field, $v_g$.     %{\bf more on this?}

If we expand the angular dependence of $\Phi$ and $\delta_0(\vs)$
redshift space   in spherical harmonics in the form, 
\begin{equation}
\Phi(\vs)=
\sum_{l=0}^{\infty}\sum_{m=-l}^l \Phi_{lm}(s)Y_{lm}(\theta,\varphi)
\end{equation}
and similarly for $\delta_0$, then, to first order,
$\Phi_{lm}$ and $\delta_{0lm}$ satisfy,
\begin{eqnarray}
\label{eq:phis}
{1\over {s^2}} {\dd \over {\dd s}}
\left(s^2 {{\dd \Phi_{lm}} \over {\dd s}}\right)
&-&{1 \over {1+\beta}}{{l(l+1) \Phi_{lm}} \over {s^2}}\\
&=&{\beta \over {1+\beta}} \left(\delta_{0lm} - 
{ \kappa(s) } { {\dd \Phi_{lm}} \over {\dd s}}\right)\; , \nonumber
\end{eqnarray}
where 
\begin{equation}
\kappa=\frac {\dd \ln\phi}  {\dd  s} - \frac{2}{s} \frac{d \ln w(L_{0i})}{d \ln L_{0i}} 
\end{equation}
represents the  correction for the bias introduced by 
 the generalized Kaiser rocket effect.  As emphasized by
ND94, the solutions  to equation (\ref{eq:phis}) for the monopole ($l=0$) and the dipole
($l=1$) components of the radial peculiar velocity in the LG frame 
are uniquely determined by specifying vanishing velocity at the origin.  
That is, the radial velocity field at redshift $\bf s$,  when expanded to
harmonic $l \le 1$, is not influenced by material at redshifts greater
than $\bf s$.  

In this paper, we shall consider solutions as a function 
of $\beta$.  \cite{Davis11} also fit a  second  parameter, $\alpha$, defining a power law form $w_i\propto L_i^\alpha$ for
the galaxy weights and found  $\alpha \approx 0$ was the best fit. The large-scale gravity field is best estimated if all the galaxies are equal-weighted, that is, they all have the same mass.   This makes sense if you remember that each point in the 2MRS represents the mass on scales of 
$\sim 4$ Mpc.

\section{Generating Peculiar Velocities}
Given a sample of galaxies with measured circular velocity
parameters, $\eta_i  \equiv {\rm log}\omega_i$, linewidth $\omega_i$, apparent magnitudes $m_i$, 
and redshifts $z_i$, the goal is to derive an estimate for the smooth
underlying peculiar velocity field.  We assume that the circular
velocity parameter, $\eta$, of a galaxy is, up to a random
scatter, related to its absolute magnitude, $M$, by means of a
linear {\it  inverse} Tully-Fisher (ITF) relation, i.e.,
\begin{equation}
\label{eq:ITF}
\eta=\gamma M + \eta_0 .  
\end{equation}

One of the main advantages of inverse TF methods is that  samples selected by magnitude, as most are, will be minimally plagued by   Malmquist bias effects when analyzed in the inverse direction  \citep{schechter, a82}. 
We write the absolute magnitude of a galaxy, 
\begin{equation}
M_i = M_{0i} + P_i
\end{equation}
where
\begin{equation}
M_{0i} = m_i + 5{\rm log}(z_i)-15 
\end{equation}
and 
\begin{equation}
P_i = 5{\rm log}(1- u_i/z_i)
\label{eq:P_i}
\end{equation} 
where $m_i$ is the
apparent magnitude of the galaxy, $z_i$ is its redshift 
in units of $\kms,$
and $u_i$ its radial peculiar velocity in the LG frame.

\section{The Solution in Orthogonal Polynomials}
Functions based on $Y_{lm}$  are a poor description.of the complex flows of LSS,  giving rise to correlated residuals, but with only $\sim 20$ numbers to describe the field, we  get    $\chi^2 /{\rm dof} = 1$ when we compare the gravity and velocity fields; 
 25 years ago, the same comparison gave      $\chi^2 /{\rm dof} = 2$ \citep{DNW96}.  In the interval,  the {\it IRAS} gravity field has been replaced by the 2MRS, but the two gravity fields are essentially identical. 
 The TF data has been updated to the SFI++ catalogue, which makes all the difference; the old data was constructed of 4 separate catalogues and it was not uniformly calibrated.
 
The choice of radial basis functions for the expansion of the modes can be made with considerable latitude. The functions should obviously be linearly independent, and close to orthogonal when integrated over volume. They should be smooth and close to a complete set of functions up to a given resolution limit.   Spherical harmonics and radial Bessel functions are an obvious choice, but Bessel functions have a constant radial resolution with  distance whereas the measured peculiar velocities have  velocity error that  scales linearly with distance.   
We deal with this problem by  choosing  to make the Bessel functions a function of $y$ instead of $r$  by means of the transformation
\begin{equation}
y(r)  = ({\rm log}(1 +  (r/r_s))^{1/2}
\end{equation}
where $r_s = 5000$ km/s.  The resulting radial functions oscillate more rapidly toward the origin then they do toward the outer limit,  a physically desirable behavior.

%\section{Fitting Mode Coefficients to the Fields}
%
%?2  behavior clearly favors low values of ?.  Best value is ?=0.33 ± 0.04
%The noise is limited by the TF scatter, not the 2MRS reconstruction.  

%With ?min= 0.33 and f(½)=.483, we get a bias factor b=1.46 ± 0.20 between the dark matter and the 2MRS galaxy distribution.  ?8  = 0.66 ± 0.10  (1.5 ? from WMAP result)

%Favor no correlation between M and mass.

% We find no evidence for SUPER  large-scale flows. We see no evidence that the dipole in the CMBR is produced by anything other than our motion in the universe.

\section{The Resulting Velocity and Gravity Fields}

\begin{figure*}
\centering
%\epsfile{file=radial.modes.eps,scale=0.8}
\includegraphics[ scale=0.6,angle=00]{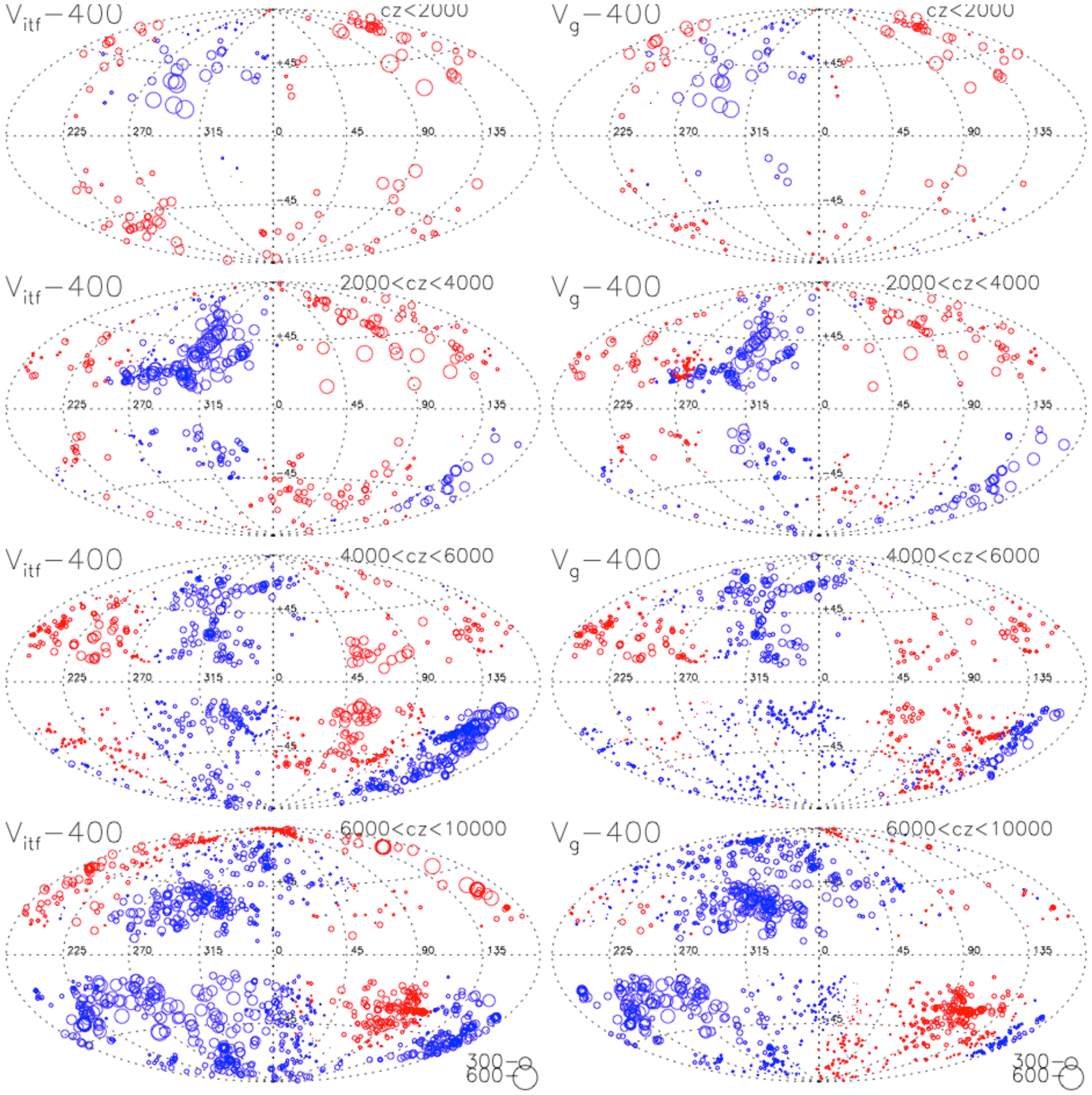}
\vspace{0pt}
\caption{  The derived peculiar velocities ${\bf \vitf}$ and $\vg$  of SFI++ galaxies on aitoff projections on the sky in galactic 
coordinates. Red points have positive peculiar velocity and blue points have negative.  The rows correspond to galaxies with $cz<2000$, $2000<cz<4000$ 
$4000<cz<6000\kms$ and $6000<cz<10000$ km/s, respectively. The size  of the symbols is linearly proportional to the velocity amplitude (see key to the size of the symbols given at the bottom of the figure). In order to better see the differences,  a 400 km/s dipole, in the direction of the CMB dipole, has been subtracted from the ${\bf \vitf}$ and $\vg$ velocities. }
\label{fig:vitf}
\end{figure*}

In the aitoff projections in Figure 1 we plot the TF peculiar velocities of the SFI++ galaxies, $\vvitf$ and the derived gravity modes, $\vg$ , for galaxies in redshift shells, $cz < 2000$, $2000 < cz < 4000$, $4000 < cz < 6000$, and $6000 < cz < 10000$ km$s^{-1}$. The projections are in galactic coordinates centered on l,b = 0 and with $b=90$ at the  top. Figure 1 is shown with $\beta=0.35$; the amplitude of $\vg$ is almost linear with $\beta$, giving a powerful diagnostic.  Our best fit is $\beta =  0.33 \pm 0.04$.  The key point is to note that the residuals are small for the entire sky and have amplitude that is constant with redshift. The amplitude and coherence of the residuals ${\bf \vitf}  - \vg$ is the same as for the mock catalogs in figure 2, where for example the lower picture shows 
${\bf \vitf } - \vg$ for real and mock catalogs. The mocks show the viability of the full procedure \citep{Davis11}.
%It is not very dissimilar from the real plot of${\bf  vitf ? vg}$ in the upper panel, demonstrating the feasibility of the entire method.

Figure 1 says it all -- the agreement between the inferred velocity field and the gravitational expectations is spectacularly good at all distances. These two fields could have been very discrepant; the only parameter of the fitting is $\beta$.  The flow field is complex, as galaxies respond to their local gravity field. All the argumentation of 25 years ago is irrelevant.  Note that we are only using 20 numbers to describe the local field, thus smoothing out the small scale velocity field.

\subsection{Residual Velocity Correlations}

\begin{figure}
\centering
%\epsfile{file=radial.modes,scale=0.8}
\includegraphics[ scale=.6,angle=00]{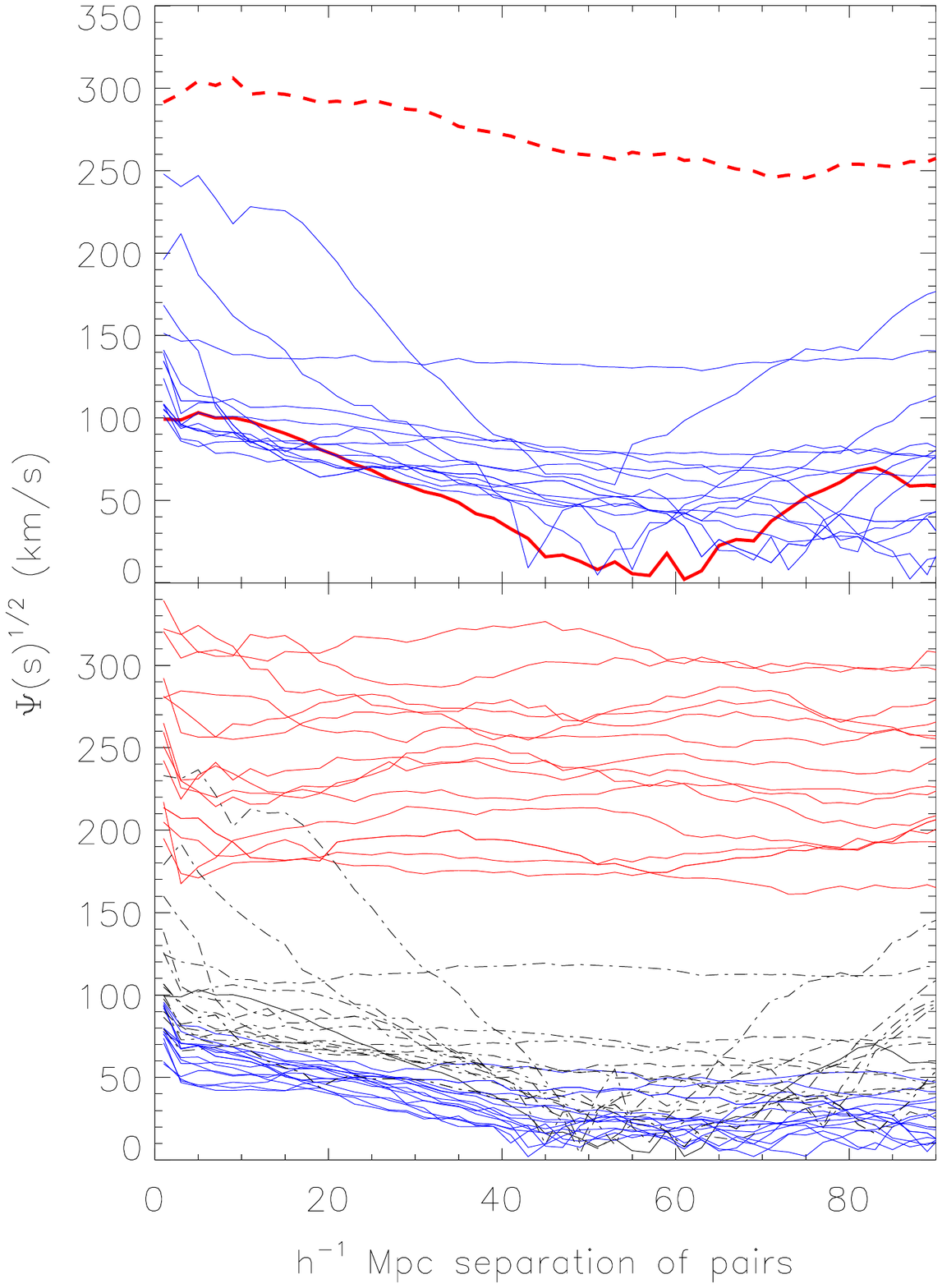}
\vspace{10pt}
\caption{{\it Top: } The velocity correlation of the real data and 15 mock catalogs. 
The dashed red and solid red curves curve are, respectively,   the correlations of $ \vitf$ and ${\bf \vitf} - \vg$ in  the real data. This plot shows that the 20 mode expansion removes virtually the entire velocity field.
  The blue lines are each correlations of $\vitf - \vg$ for the mock catalogs.  {\it Bottom:}  Velocity correlations  for 15 mock catalogs.  The red curves are
   the velocity of ${\bf v_{itf}}$, the dot-dashed curves show the correlation of ${\bf (v_{true}-v_{g})}$, and the blue curves correspond to  
   ${\bf v_{true}-v_{itf})}$.  
Both ${\bf v_{true}}$ and ${\bf v_{g}}$ are first smoothed with the 20 mode expansion before the autocovariance is computed. Note that the correlation of 
${\bf \vitf}-\vg$ is only slightly worse than the correlation of ${\bf v_{true}-v_{gs}}$, showing that the velocity reconstruction dominates the errors.  Note also that we are plotting the square root of the velocity correlation $\Psi$.  }
\label{fig:psiv}
\end{figure}

The residuals, both in the real and mock data, have error fields, ${\bf \vitf} - \vg$,  that show large regions of coherence.   To address the significance of these errors, we show in figure 2 the velocity correlation function \citep{gd89} defined as 
\begin{equation}
\label{eq:Psiv}
\Psi(s; {u})=\frac{\sum_{pairs}  u_1u_2 {\rm cos}\theta_{12}}{\sum_{pairs} {\rm cos}^2\theta_{12}}
\end{equation} 
where  the sum is over all pairs, 1 and 2, separated by vector distance $\bf s_{12}$  
(in redshift space),  $ {\rm \theta_{12}}$ is the angle between points $1$ and $2$,  and $u$ is either $\vitf$ 
(dashed red) 
or ${\bf \vitf} - \vg$ (red for data, blue for 15 mock catalogs),    
At small lags for the real data, the function $\Psi(r; {\bf \vitf}-\vg)$ is  a factor of 3 less than 
$\Psi(s;{\bf \vitf})$, about the same as for the mock catalogs. Note how the large coherence of ${\bf \vitf}$ is enormously diminished in 
$\Psi(s>2000 {\rm km/s};{\bf \vitf}-\vg)$. This shows that the coherence seen in the residual field, 
figure \ref{fig:psiv}, is expected and is not a problem. The large scale drift of a sample is demonstrated by the persistent amplitude of $\Psi$ beyond $\approx 60-80$ Mpc.

The bottom panel of figure \ref{fig:psiv} shows velocity correlations for 15 mock catalogs where the actual velocity ${\bf v_{true}}$ generated in the nbody code and then smoothed with the 20 mode expansion can be compared to either ${\bf \vitf}$ or $\vg$.   Note that the raw velocities, ${\bf v_{itf}}$ (red), have enormous correlation that reaches large lag, while the correlations, $({\bf v_{true}-\vitf})$, (blue) are extremely small.  This is because the only difference with $v_{true}$ is the gaussian error in $\Delta\eta = .05$ that affects ${\bf \vitf}$.  The blue curves show this error is not a problem, because the mode expansions are insensitive to gaussian noise in the 2500 galaxies, i.e. they are essentially perfect.  This demonstrates that even though the TF noise is as large as for the actual data, the ability to find the correct flow, when characterized by only 20 numbers, is intact.  

This demonstrates that the description of the full velocity field by the specification of 20 numbers, specifying the amplitude of the modes, is essentially complete.

\section{Summary}

\begin{itemize}
\item{ We see no evidence that the dark matter does not follow the galaxy distribution,  and it is consistent with constant bias on large scales. There is no evidence for a non-linear bias in the local flows. A smooth component to the universe is not something testable with these methods.}

\item{ Linear perturbation theory appears to be adequate for  the large scales  tested by our method; the comparison of ${\bf v_p}$ and ${\bf g}$ is so precise as to be a stunning example of the power of linear theory! }
 
\item{Our estimate of $\sigma_8 $ gives the most precise value at $z\sim 0$ and is useful for tests of the growth rate and Dark Energy.}

\item{The velocity-gravity comparison measures the acceleration on scales in the range $10 - 60$ Mpc. and  since we derived a  similar value of $\beta$  as for clusters of galaxies, we conclude  that dark matter appears to fully participate in the clustering on scales of a few Megaparsecs and larger.} 

\item{We find no evidence for large-scale flows, and the small residuals are completely consistent with LCDM \citep{2014ApJ...788..157N}.  Note that our analysis has not used the CMBR dipole,  but we see a velocity field that is fully consistent with the  CMBR  dipole radiation. We see no evidence that the dipole in the CMBR is produced by anything other than our motion in the universe.}

\item{The field of Large Scale Flows, apart from going deeper with TF data, appears to this observer to have finally reached its  original goal.  Remember that 25 years ago, there were no CMBR results measuring $\Omega_m$, and the large scale flows were going to give us the long-sought answer.  But the TF data of 25 years ago was not well calibrated and gave inconsistent results, so we lost ground.  Now we can state that the LS flows are consistent with standard parameters. }

\item{This finishes the study of the local velocity field, and now I can retire!}

\end{itemize}

%\bibliographystyle{astron}
%\begin{thebibliography}{}
%\bibliography{References}{}

%\end{thebibliography}

\end{document}